\documentclass[aps,prl,twocolumn,showpacs]{revtex4}
\usepackage{amssymb}
\usepackage{graphicx}
\usepackage{epstopdf}
\usepackage{amsmath}
\usepackage{epsfig}
\usepackage{color}

\begin{document}
\title{
Fractional lattice charge transport }
\author{Sergej Flach$^{1,2}$, Ramaz Khomeriki$^{1,3}$, }
\affiliation {${\ }^1$Center for Theoretical Physics of Complex
Systems, Institute for Basic Science, Daejeon, South Korea \\
${\ }^2$New Zealand Institute for Advanced Study, Centre for
Theoretical Chemistry \& Physics, Massey University, Auckland, New
Zealand \\ ${\ }^3$Physics Department, Tbilisi State University,
Chavchavadze 3, 0128 Tbilisi, Georgia }
\begin{abstract}
We consider the dynamics of noninteracting electrons on a square
lattice in the presence of a magnetic flux $\alpha$ and a dc
electric field $E$ oriented along the lattice diagonal. In general,
the adiabatic dynamics of an electron will be characterized by Bloch
oscillations in the electrical field direction and dispersive ballistic
transport in the perpendicular direction. For rational
values of $\alpha$ and a corresponding discrete set of values of
$E(\alpha)$ vanishing gaps in the spectrum induce a
fractionalization of the charge in the perpendicular direction -  while left movers are still performing
dispersive ballistic transport, the complementary fraction of right movers is propagating in a
dispersionless relativistic manner in the opposite direction. Generalizations to
other field directions, lattice symmetries and the possible probing of the effect with atomic Bose-Einstein condensates and photonic networks are discussed.
\end{abstract}
\maketitle

The two-dimensional electron gas in a perpendicular magnetic field
is a celebrated topic in condensed matter physics (see e.g. Ref.
\cite{girvin}). Electron-electron correlations or electron-lattice
interactions lead to fractional quantum Hall physics, while the
integer quantum Hall effect is based on the properties of the single
particle eigenstates in the presence of a weak dc electric field in
the linear response regime. Underlying discrete lattice structures
and symmetries can have substantial impact on the wavefunctions. The
well-known case of a two-dimensional square lattice leads to the
much-studied Harper model \cite{harper} by reducing the
two-dimensional problem to the dynamics in a one-dimensional
quasi-periodic potential. Interestingly the interplay of the
two-dimensional lattice structure with magnetic fields {\sl and} a
substantial in-plane electric field is far from being well
understood, despite some notable publications \cite{jpn,nazareno}
and \cite{kol1}, where asymmetric spreading regimes have been
observed for few different directions of the in-plane electric field
with respect to the lattice axes \cite{kol2}.
\begin{figure}[b]
\includegraphics[width=0.9\columnwidth]{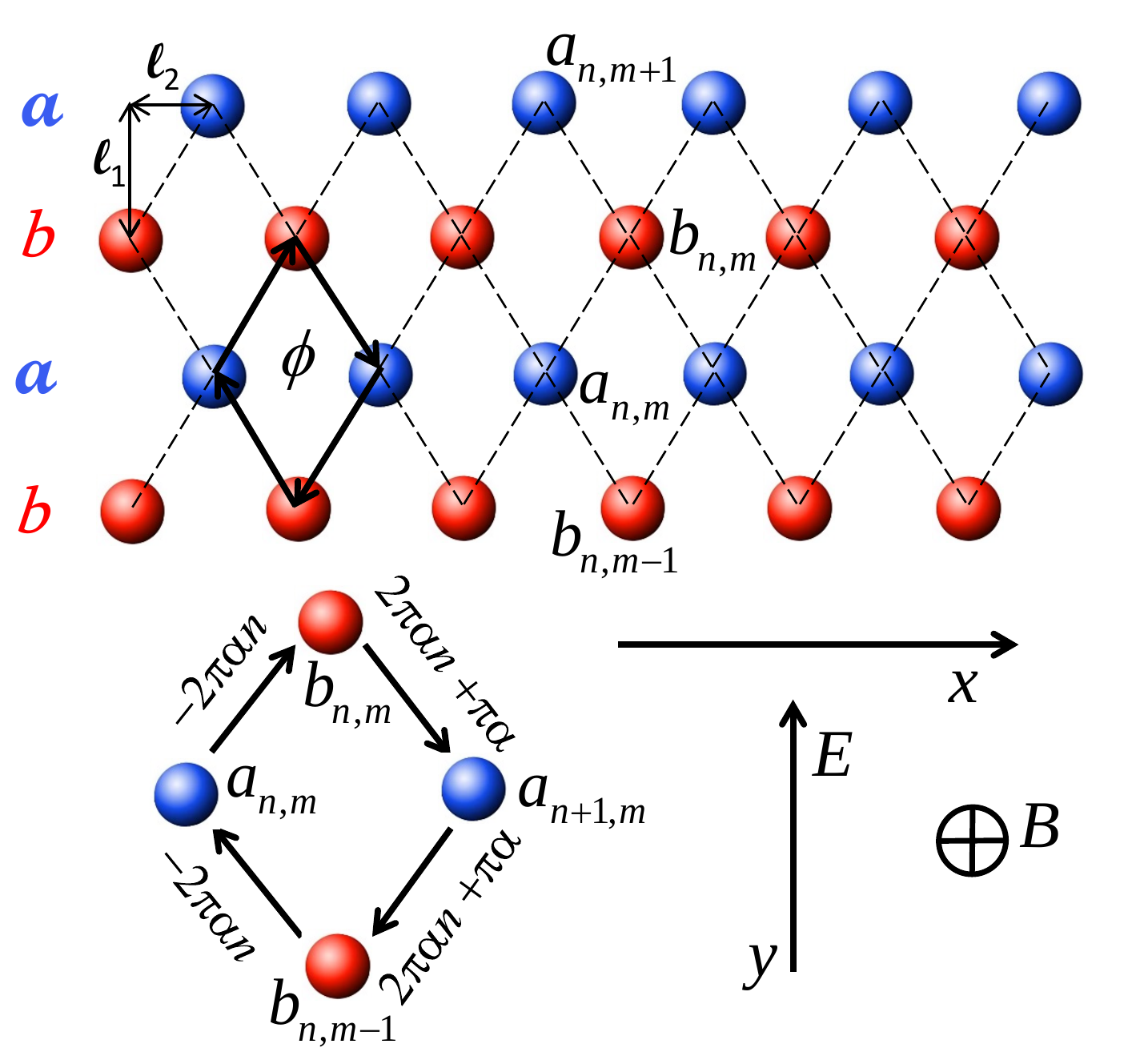}
\caption{Schematics for the square lattice with legs $a$ (blue) and $b$ (red).
Dashed lines connect sites with allowed hopping/tunneling of the particle. The
magnetic flux $\phi$ is induced by the perpendicular magnetic field
$B$ and traverses the elementary rhombus with half diagonals
$\ell_1$ and $\ell_2$. The corresponding phases $\gamma_{ll'}$ are
shown in the lower zoom of one plaquette, and the arrows indicate
the direction of  integration of (\ref{101}). The dc electric field
$E$ is oriented along the $y$-axis. Therefore charge transport is
observed along the $x$ direction.} \label{fig1}
\end{figure}

While the generalized translational invariance - with shifts in
space and energy - is preserved in the presence of an electric
field, details do depend on the orientation. A field orientation
along a main lattice axis is the most simple yet trivial case since
the shift is identical with the lattice spacing. A general
orientation angle of the electric field relative to the lattice axes
will lead to potentially very large or possibly even infinite
shifts. We investigate the nontrivial case with the electric field
being oriented along the diagonal of the square lattice, which leads
to a period doubling of the shifts. The band structure is in general
given by infinitely many interconnected bands gapped away from each
other.
We show here that the gaps between the bands
surprisingly vanish at particular values of the electric field.
Treatable cases correspond to rational relative magnetic flux
values. The resulting band structure is given by intersecting left and right mover bands
with opposite average group velocities.
Moreover, the unexpected outcome is  that the right movers
have a relativistic linear dispersion.
Wavepackets of initially localized particles are
then shown to split into two parts, with a fractional relativistic
current of right movers. While
experiments using a two-dimensional electron gas might be a
challenging task, our results could be directly verified in the
context of Bose-Einstein condensates in optical lattices where the
effective electric field is generated by a tilt of the lattice in
the gravitational field \cite{kasevich} or accelerating a whole
lattice \cite{camp}, while the magnetic field is produced by
artificial gauge fields \cite{gauge}. Further, light propagation in
waveguide networks can emulate the electric field analogy with a
curved geometry of the waveguides \cite{longhi}, while a special
metallic fabrication of the waveguides and the surrounding medium
\cite{metal} leads to phase shifts of tunneling rates which leads to
a magnetif field analogy.

We consider the following Hamiltonian describing single electron dynamics in
a tight-binding lattice (see Fig. \ref{fig1}), in the presence of out-off plane
magnetic $B\parallel z$ and in-plane electric $E\parallel y$ dc
fields:
\begin{eqnarray}
{\cal \hat H}=\sum\limits_{\langle
\ell\ell^\prime\rangle}e^{i\gamma_{\ell\ell^\prime}}\hat
q_\ell^+\hat q_{\ell^\prime}+\sum\limits_{\ell}\ell_yE\hat
q_{\ell}^+\hat q_\ell \;. \label{1}
\end{eqnarray}
Here $\hat q_{\ell}^+$ and $\hat q_{\ell}$ are standard particle
creation and annihilation operators at the $\ell$-th lattice site.
The voltage drop between neighbouring $a,b$ chains is denoted by $V$.
The first
term in the Hamiltonian accounts for the hopping between nearest neighbor
sites $\ell$ and $\ell^\prime$ in the presence of a magnetic field.
The corresponding
phase factor is given \cite{vidal}:
\begin{eqnarray}
\gamma_{\ell\ell^\prime}=\frac{2\pi
e}{hc}\int_\ell^{\ell^\prime}\vec A d\vec s\;. \label{101}
\end{eqnarray}

The vector potential is defined in the Landau gauge as $\vec A\equiv
(0,~xB,~0)$. We denote the flux through the elementary rhombus (with
diagonals 2$\ell_1$ and 2$\ell_2$) as $\phi=2B\ell_1\ell_2$. With
the flux quantum $\phi_0=hc/e$, the relative flux is defined as
$\alpha=\phi/\phi_0$. The voltage drop between $a,b$ chains follows
as $V=El_1$. The Schr\"odinger equation
\begin{equation}
i\partial \Psi / \partial t = {\cal \hat{H}} \Psi
\label{erwin}
\end{equation}
 describes the evolution of the particle wave function $\Psi$
with time-dependent complex probability amplitudes $a_{n,m}$, $b_{n,m}$ assigned to
all lattice sites. With the notations $c_{n,2m}=a_{n,m}$,
$c_{n,2m+1}=b_{n,m}$, $\theta = 2\pi \alpha$ (flux angle) and the transformation
\begin{eqnarray}
c_{n,l} = C_l e^{i(kn-\lambda t)} e^{-i\pi \alpha n(2l-1)}
\label{trans}
\end{eqnarray}
which takes the space direction $x$ (index $n$) transversal to the
applied electrical field into Fourier space with wave number $k$,
we arrive at a simple
one-dimensional bipartite chain equation (see \cite{supp} for derivation details)
\begin{eqnarray}
&\lambda {C}_{l} =(l-\frac{1}{2})V C_{l}- [1+e^{(-1)^l i(-k +
l\theta) }]C_{l+1} \nonumber \\
&- [1+e^{(-1)^l i(-k + (l-1)\theta)}]C_{l-1} \;. \label{eqc}
\end{eqnarray}
\begin{figure}[b]
\includegraphics[scale=.75]{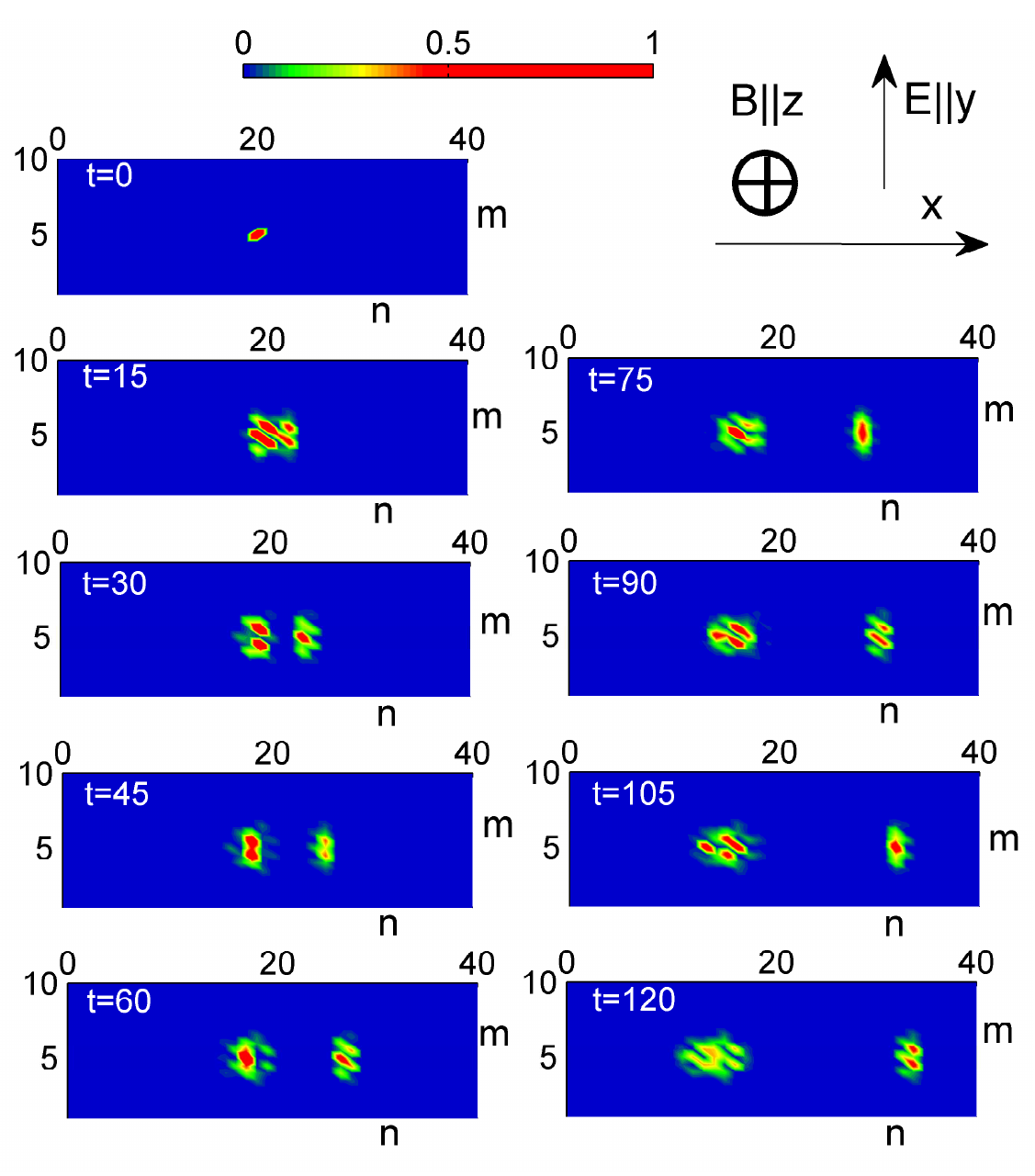}
\caption{The evolution of the wave packet ($|a_{n,m}|^2$ and
$|b_{n,m}|^2$) of a single electron placed initially in the center
of the two-dimensional lattice.  Distributions for different times
are presented. The right top graph shows the  orientation of the dc
electric and magnetic fields.} \label{fig2}
\end{figure}
For each value of $k$ the equation set (\ref{eqc}) corresponds to a
generalized Wannier-Stark ladder with a discrete unbounded spectrum
$\lambda_{\nu}(k)$. Bands with different indices $\nu$ will
generically avoid intersections upon varying the wave number $k$ due
to level repulsion \cite{neumann-wigner}. As a result, an initially localized electron
will be trapped and perform generalized Bloch oscillations in the $y$-direction.
At the same time, on ballistic dispersive spreading will occur in the $x$-direction due to the overlap of the initial
state with states from an effective finite number of bands.
With $\partial ^2
\lambda_{\nu}(k) / \partial k^2 \neq 0$ the spreading will be dispersive in both $x$-directions,
since a whole spectrum of group velocities will lead to a widening of the wave packet.

We do observe this outcome in general, however, we find that for
each value of relative magnetic flux $\alpha$ a value of the
electric field exists, for which a localized fraction of the wave
packet is propagating in the {\sl opposite direction} with a well
defined velocity $V/(2\pi\alpha)$.
We also observe that precisely for those parameter values the gaps in the above discussed band structure
vanish (this can indeed happen for matrices with elements depending on more than one parameter \cite{neumann-wigner}).
In Fig. \ref{fig2} the wave
packet is shown for $\alpha=1/3$ and $V=\sqrt{4.8}$ at different
times. Indeed one third of the wave packet is propagating in a
nondispersive localized manner to the right, while the complementary
wave packet part is spreading as usual to the left (the wave packet
dynamics is obtained by integrating equation (\ref{erwin})
in time).

Clearly the condition for the occurence of a charge
fractionalization must be routed in the zeroing of gaps in the band
structure, which lead to effective left- and right-movers. To
proceed we investigate the matrix $\hat{M}$ whose zero determinant is yielding
the eigenvalues $\lambda$ of (\ref{eqc}):
\begin{widetext}
\begin{eqnarray}
\hat M=\left(
  \begin{array}{cccccccccc}
    \ddots & \vdots & \vdots & \vdots & \vdots & \vdots & \vdots & \vdots & \vdots \\
    \cdots & 0 & -1-e^{i(k+2\theta)} & -\lambda-\frac{3V}{2} & -1-e^{i(k+\theta)} & 0 & 0 & 0 & 0 &  \cdots\\
    \cdots & 0 & 0 & -1-e^{-i(k+\theta)} & -\lambda-\frac{V}{2} & -1-e^{-ik} & 0 & 0 & 0 & \cdots\\
    \cdots & 0 & 0 & 0 & -1-e^{ik} & -\lambda+\frac{V}{2} & -1-e^{i(k-\theta)} & 0 & 0 & \cdots\\
    \cdots & 0 & 0 & 0 & 0 & -1-e^{-i(k-\theta)} & -\lambda+\frac{3V}{2} & -1-e^{-i(k-2\theta)} & 0 & \cdots\\
     & \vdots & \vdots & \vdots & \vdots & \vdots & \vdots & \vdots & \vdots & \ddots\\
  \end{array}
\right) \label{41}
\end{eqnarray}
\end{widetext}

The band structure $\lambda_{\nu}(k)$ follows from
$Det[\hat M]=0$.
Matrix (\ref{41}) is invariant under the symmetry operation
\begin{equation}
\lambda \rightarrow \lambda-V\;, \; k \rightarrow k+\theta  \;.
\label{symmetry}
\end{equation}
Further, the spectrum $\lambda_{\nu}(k)$ is invariant under the symmetry operation
\begin{equation}
k \rightarrow -k \;,\; V \rightarrow -V \;.
\end{equation}
It follows, that if the eigenenergy $\lambda$ is degenerated for some values of the voltage $V$ and wavenumber $k$,
then the eigenenergy $\lambda^\prime=\lambda-V$ exists and is also degenerated
for the same voltage $V$ at wavenumber $k^\prime=k+\theta$.
Therefore closing one gap in the spectrum implies closing all symmetry related gaps as well.
For the particular case of $\lambda=0$ and $k=\pi$ the matrix $\hat{M}$ splits into two semi-infinite blocks, and we then arrive at the general statement that
gaps must close for particular values of pairs of $\theta$ and $V$ (see \cite{supp} for details).

\begin{figure}[h]
\includegraphics[scale=.68]{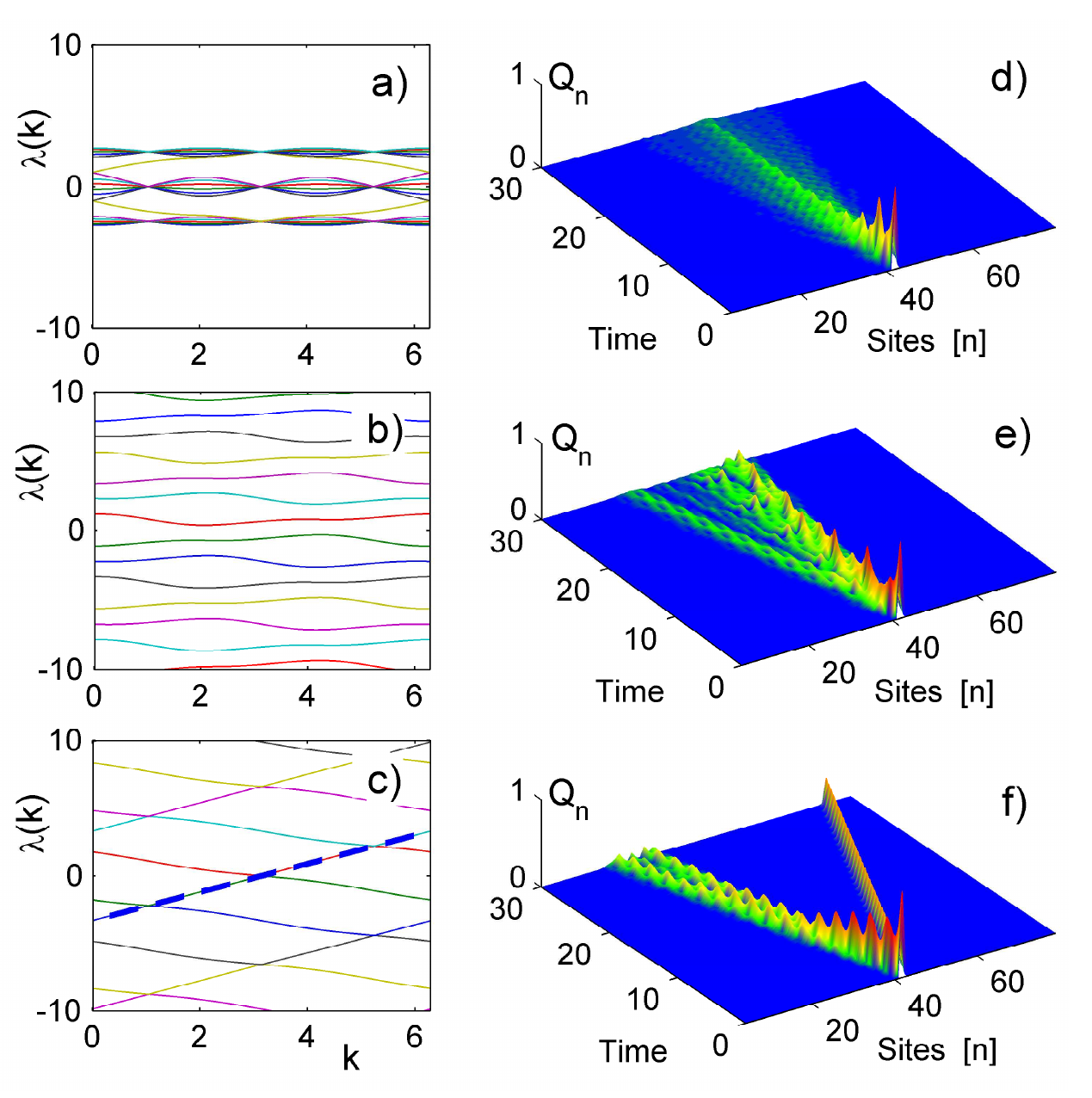}
\caption{Graphs a)-c): Dispersion relations for the Bloch bands
after diagonalizing matrix \eqref{41} with dimensions $80 \times 80$.The  relative flux is
$\alpha=1/3$ and the voltage drop take the values (from top to
bottom) $V=0$, $V=1$ and $V=\sqrt{4.8}$.
Graphs d)-f) show the wavepacket spreading upon integrating
\eqref{erwin} with a single site initial condition using the parameters
of the respective left panel graphs. The dynamics of the integrated
charge density \eqref{Q} accumulated in the $n$-th cross section is
presented. The dashed line in graph c) corresponds to the
dispersionless curve $\lambda(k)=\pi+2\pi k/3$.} \label{fig3}
\end{figure}

Rigorous results are obtained for
rational values of $\alpha$.
In these cases the matrix $\hat{M}$ splits into noninteracting block matrices of finite size.
Consider $\alpha=1/3$. The band structure is shown in Fig.\ref{fig3}(a-c) for various voltage drops.
The matrix $\hat{M}$ splits into $3\times 3$ block matrixes, with one of them given by
\begin{eqnarray}
\hat M^{(3)}=\left(
  \begin{array}{ccc}
    \frac{V}{2} & 1-e^{-2i\pi/3} & 0 \\
    1-e^{2i\pi/3} & \frac{3V}{2} & 1-e^{4i\pi/3} \\
    0 & 1-e^{-4i\pi/3} & \frac{5V}{2}\\
  \end{array}
\right) . \label{403}
\end{eqnarray}
The condition $Det\left(\hat M^{(3)}\right)=0$ yields the roots
$V=0$ and $V=\pm\sqrt{4.8}$.
It follows that
for relative flux $\alpha=1/3$ and voltage drop $V=\pm\sqrt{4.8}$ a
nontrivial gap closing takes place which is indeed observed in Fig.
\ref{fig3} c).
We then consider a particle initially localized on one site
evolve this state in time according to (\ref{erwin}).
We compute the integrated
charge density
\begin{equation}
Q_n=\sum\limits_l|c_{n,l}|^2 \label{Q}
\end{equation}
and plot the result in Fig. \ref{fig3}(d-f). While the cases $V=0,1$ yield the typical
dispersive wave packet spreading, we find that for $V=\sqrt{4.8}$ exactly one third of the wave packet
is propagating in a relativistic nondispersive manner to the right, leading to the effect of charge separation
(see \cite{supp} for a detailed calculation of the charge separation value).

For general rational relative flux values
$\alpha=p/s$ one has to consider an $s\times s$ matrix $\hat
M^{(s)}_{\alpha=p/s}$. Then the condition
$Det\left(M^{(s)}_{\alpha=p/s}\right)=0$ gives $s$ real roots for $V=\pm |V|$.

\begin{figure}[t]
\includegraphics[scale=.56]{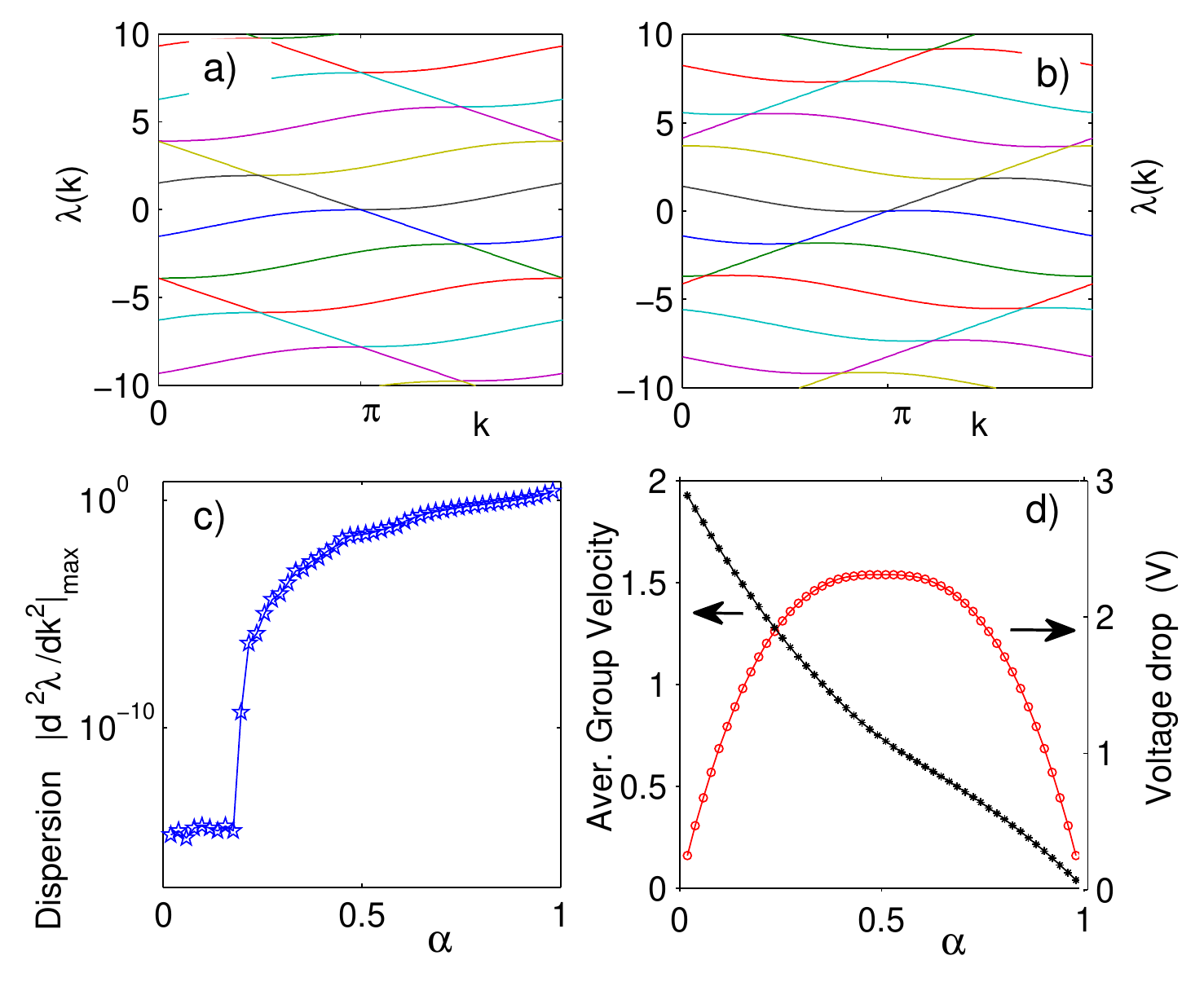}
\caption{ Upper plots: Band structure for relative flux values
$\alpha=3/4$, $\alpha=2/9$ and voltage drops $V=1.8322$ (a) and
$V=0.8676$ (b). Lower plots: The $\alpha$ dependence of the largest
dispersion $|\partial^2 \lambda / \partial k^2|$ (c), of the
averaged group velocity $|\langle \partial \lambda / \partial k|$
(black stars) and of the largest voltage drop value $V_1$ (red
circles)  (d) for the analyzed cases of charge separation in the gap
closing regime. Note that only rational values of $\alpha$ are
considered here.
 } \label{fig4}
\end{figure}
The root with the largest absolute value will yield $p/s$ fractional
transport, which can be found either analytically or numerically. In
the upper panel of Fig. \ref{fig4} band structures for two different
relative flux values $\alpha=3/4$ and $\alpha=2/9$ are shown in the
gap closing regime ($V=1.8322$ and $V=0.8676$ respectively).
Fractional transport with exactly $3/4$ and $2/9$, respectively, of
the total charge is observed in numerical simulations. In the lower
panel of Fig.\ref{fig4} we plot the  largest dispersion $|\partial^2
\lambda / \partial k^2|$ (c), and the averaged group velocity
$|\langle \partial \lambda / \partial k|$ (d) for the analyzed cases
of charge separation in the gap closing regime. While the dispersion
is very small but nonzero for $\alpha \geq 1/6$, it vanishes for
$\alpha < 1/6$. In the same limit of small $\alpha$ values the group
velocities of the fractional charge tend to their largest values.

So far we considered charge separation for the largest root of
voltage drop given by $Det\left(M^{(s)}_{\alpha=p/s}\right)=0$.
Interestingly, the other roots yield an even more complex charge
separation scenario. For $\alpha=1/11$ we examine the corresponding
block matrix $M_{\alpha=1/11}^{(11)}$ with dimension $11\times 11$.
The gaps close when the condition
$Det\left(M_{\alpha=1/11}^{(11)}\right)=0$ is fulfilled which
produces five nontrivial independent positive roots: $V_5=0.0869$,
$V_4=0.2244$, $V_3=0.4135$, $V_2=0.6585$ and $V_1=0.9660$.
\begin{widetext}
~
\begin{figure}[b]
\includegraphics[scale=.85]{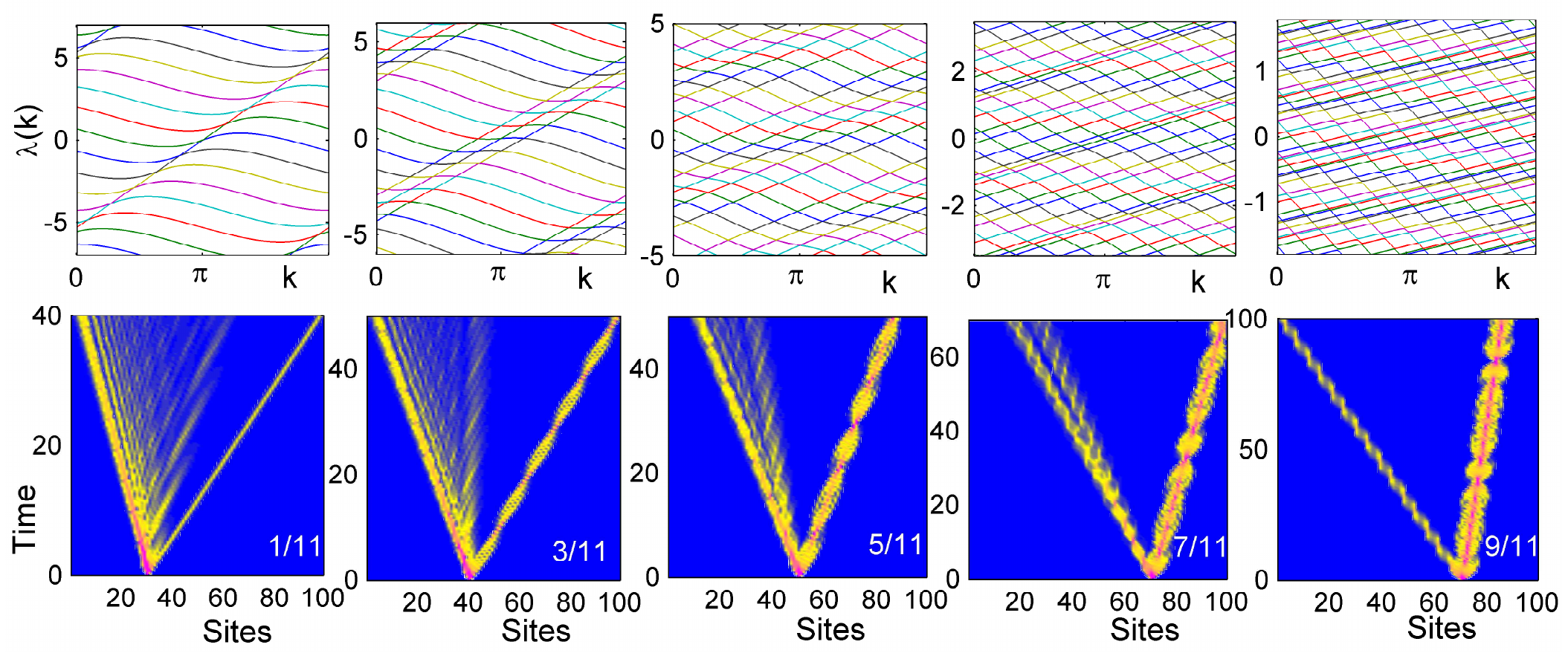}
\caption{Same as in Fig. \ref{fig3} but with different parameters:
the relative magnetic field flux is $\alpha=1/11$ and the voltage
drop values are the five roots of the equation $Det\left(\hat
M_{\alpha=1/11}^{(11)}\right)=0$, in particular
$V=0.9660,~0.6585,~0.4135,~0.2244,~0.0869$ (from left to right).
Note different time scaling in lower panel. Corresponding video
files representing charge dynamics in 2D lattice could be found in
supplemental material. } \label{fig5}
\end{figure}
~
\end{widetext}

The corresponding band structure (upper panel) and fractional charge
dynamics (lower panel) with single site initial conditions are shown
in Fig. \ref{fig5}. For the largest root $V=V_1$  a fraction of
$1/11$ of the charge is separated and is propagating
relativistically (see \cite{supp} for numerical evidence). This
follows straight from the band structure, since only one of the
eleven bands is yielding a positive nondispersive group velocity
(see first column of Fig. \ref{fig5}). The dependence of this
largest voltage drop $V_1$ on various rational values of $\alpha$ is
shown in Fig.\ref{fig4} (d). We observe that in the limit of weak
magnetic fields the corresponding voltage drop values tend to zero
as well, however keeping the surprising feature of fractional
relativistic transport.

For the other roots more bands are contributing to the fractional
current, while still only one is completely gapless, as shown in
Fig.\ref{fig5}. Moreover, the nonzero gaps become much smaller,
leading to almost relativistic charge separation in both directions
(lower panels in Fig.\ref{fig5}). The value of the charge fraction
is now given by the number of the participating bands times the
value of $\alpha$. In our case of $V_m$ nontrivial positive roots
with $m=1,2,3,4,5$ we find that the charge fraction is given by
$(2m-1)\alpha$ (see \cite{supp} for numerical evidence).

The observed fractional charge transport is crucially linked to the orientation of the electric field. In our case it is directed along the diagonal of a square lattice.
Consider instead an electric field orientation along a main axis of the square lattice. In that case the momentum $k$ of motion perpendicular to the electric field
will completely decouple from the magnetic flux. As a result the matrix $\hat{M}$ in Eq.(\ref{41}) will not decouple anymore into two noninteracting blocks at a special
value of $k$, since $k$ will enter its diagonal part only. This will destroy the exact degeneracies which lead to fractional transport.

In summary, we demonstrate that a simple quadratic lattice is sufficient to obtain fractional charge separation of noninteracting electrons,
or ultracold atomic gases, in the presence of magnetic fields, or synthetic gauge fields and a properly oriented and tuned DC bias.
Such a charge separation can be potentially very useful for the preservation and engineering of entanglement in quantum systems.

\end{document}